\documentclass[twocolumn,prl] {revtex4}
\usepackage{graphics}
\usepackage{epsf}
\usepackage{amsmath}
\usepackage{amssymb}
\usepackage{amscd}
\usepackage{color}

\newcommand{\flabel}[1]{\label{f:#1}}
\newcommand{\elabel}[1]{\label{e:#1}}

\newcommand{\eq}[1]{Eq.~(\ref{e:#1})}

\newcommand{\fig}[1]{Fig.~\ref{f:#1}}

\newcommand{\quot}[1]{``#1''}
\newcommand{\OCAL}{\mathcal{O}}  
\newcommand{\SET}[1]{\{#1\}}

\newcommand{\sigmavec}{\boldsymbol{\sigma}}
\newcommand{\Upsilonvec}{\boldsymbol{\Upsilon}}

\newcommand{\expa}[1]{\mathrm{e}^{#1}}   
\newcommand{\expb}[1]{\exp \glb #1 \grb} 
\newcommand{\glb}{\left(}  
\newcommand{\grb}{\right)}  
\newcommand{\glc}{\left[}  
\newcommand{\grc}{\right]}  

\newcommand{\TO}{,\ldots,}
\newcommand{\dd}[1]{\text{d}{#1\ }}   

\newcommand{\mean}[1]{\left\langle #1 \right\rangle}

\newcommand{\taucoup}{\tau_{\text{coup}}}  
\newcommand{\taucorr}{\tau_{\text{corr}}}  
\newcommand{\tausim}{\tau_{\text{sim}}}  

\begin{document}

\title{Renormalization group approach to exact sampling}
\author{C\'{e}dric Chanal}
\author{Werner Krauth}
\affiliation{CNRS-Laboratoire de Physique Statistique, Ecole Normale
Sup\'{e}rieure, 24 rue Lhomond, 75231 Paris Cedex 05, France}

\date{\today}

\begin{abstract}
In this Letter, we use a general renormalization-group algorithm to
implement Propp and Wilson's \quot{coupling from the past} approach to
complex physical systems. Our algorithm follows the evolution of the
entire configuration space under the Markov chain Monte Carlo dynamics
from parts of the configurations (patches) on increasing length
scales, and it allows us to generate \quot{exact samples} of the Boltzmann
distribution, which are rigorously proven to be uncorrelated with the
initial condition.  We validate our approach in the
two-dimensional Ising spin glass on lattices of size $64 \times 64$.
\end{abstract}

\maketitle

The Markov chain Monte Carlo method \cite{metropolis} has developed
into  a universal computational approach in many disciplines of science
and engineering, and it remains of great importance in  the field of
statistical physics where it originated more than $50$ years ago. Indeed,
many difficult calculations in high-dimensional spaces can be expressed
(more or less formally) as the calculation of expectation values 
of an observable $\OCAL$
\begin{equation}
\mean{\OCAL} = \frac{\int \dd{x} \expa{- \beta E(x)} 
      \OCAL(x)}{\int \dd{x} \expa{- \beta E(x)}}
\simeq \frac{1}{\tausim} \sum_{\text{$t=t_0$}}^{t_0+\tausim}\OCAL(x_t).
\elabel{phase_space_time_average}
\end{equation}
To compute the ensemble average on the left of
\eq{phase_space_time_average}, the Markov chain [on the right of
\eq{phase_space_time_average}] passes from one configuration $x_t$ at
time $t$ to the next one, $x_{t+1}$, most often in a way respecting the
detailed balance condition.  An ergodic Markov chain, which can eventually
reach any configuration $x$ from any other $\tilde{x}$, has the property
to visit configurations with probability $\propto \expa{-\beta E(x)}$
in the limit of infinite simulation time $\tausim \rightarrow \infty$,
where the time average indeed coincides with the ensemble average.
In all Monte Carlo calculations, the convergence toward the stationary
values of means and (connected) correlation functions is exponential with,
for example,
\begin{equation}
\mean{\OCAL(x_t) \OCAL(x_{t + \tausim })}_c \simeq 
               A \expb{- \tausim /\taucorr}. 
\elabel{exponential_convergence}
\end{equation}
The correlation time $\taucorr$ provides a crucial scale, because
only Markov chain Monte Carlo simulations that have run for times much
longer that $\taucorr$ yield useful results and are essentially free of
systematic errors caused by the initial condition.

Data analysis routines allow to reliably estimate the correlation time
(and relatedly, the error of a Monte Carlo calculation) whenever $\taucorr
\ll  \tausim$ (see, for example, \cite{SMAC} for a discussion of the
\quot{bunching method}).  On the other hand, it is extremely difficult to
ascertain that a Monte Carlo simulation has indeed converged, that is,
that the essential requirement for its validity, $\tausim \gg \taucorr$
(or at least $\tausim \gtrsim \taucorr$), is satisfied. This difficulty
is very prominent, for example in the simulation of disordered systems,
where the multidimensional space of configurations $x$ is extremely
rugged, and where the Monte Carlo dynamics is governed by many different
time scales (the longest of which is the correlation time).  It is
often impossible to assure the validity of a Monte Carlo calculation in
disordered systems and other fields without resorting to the comparison
with alternative methods, like exact analytic solutions, power expansions,
careful finite-size scaling,  etc. Bhatt and Young address this point
in their classic paper \cite{Bhatt88} when discussing their convergence
test \cite{Bhatt85}:
\quot{In practice, though, one needs a criterion which determines whether
any error made in being not quite in equilibrium is acceptably small
(...). We know of no "rigorous" such criterion but we have found that
the following procedure works well in practice.}

More than a decade ago, Propp and Wilson \cite{propp}  realized a major
conceptual breakthrough: they showed how to reformulate the Markov chain
Monte Carlo algorithm so that it generates \quot{exact samples} that have
no correlation with the initial configuration, not even an exponentially
small one, as in  \eq{exponential_convergence}.  The generation of
exact samples is of greatest interest in many real-world applications:
it would solve the above-mentioned problem because the samples are exactly
in equilibrium (they carry no memory of the initial state) and because
the criterion is rigorous.

Unfortunately,  Propp and Wilson's procedure, termed \quot{coupling from
the past}, has been notoriously difficult to apply to complicated physical
systems, because it implies, as we will discuss later, the monitoring of
the entire configuration space of a physical system under the Monte Carlo
dynamics.  In the ferromagnetic Ising model above the Curie temperature,
the entire configuration space can be monitored, very elegantly, by
making use of a partial order of spin configurations (\cite{propp}, see
also \cite{SMAC,novotny}). Huber \cite{huber} has presented
an interesting method that, unlike the partial ordering approach,
can be made to work for disordered or frustrated systems \cite{Childs},
but only at very high temperature.

In the present Letter, we show how to apply exact sampling to more complex
systems than have been treated before, namely the two-dimensional
spin glass on large lattices at low temperature.  We use a general,
yet rigorous, method which monitors all the configurations in the
whole system, not directly (because there are far too many of them)
but through \quot{patches} of initially much smaller scale. During the
calculation we gradually increase the size of the patches, not unlike
what is done in renormalization-group calculations. At the latest stages
of the calculation, the patches are of the same size as the lattice,
and the configurations on patches correspond to configurations on the
entire system. Our approach to exact sampling works in a wide range of
temperatures, and for quite large systems.  We are able to judge its
efficiency by comparison with a \quot{naive} method which performs a
standard Monte Carlo simulation for a representative fraction of all
the configurations.

The two-dimensional spin glass does not have a phase transition at
temperature $1/\beta=T>0$, but it is already a quite complex system, due
to the presence of disorder. During recent years, the two-dimensional spin
glass has been a test bed for new algorithms\cite{swendsen,saul,loebl},
and has given rise to controversies concerning the specific heat capacity
at low temperature \cite{swendsen,saul,marinari}. In this model, the
convergence time of Monte Carlo calculations is difficult to estimate.

In the heat bath algorithm, the spin $\sigma_i(t)$ (on site $i=1
\TO N$) is updated using a uniform random number $\Upsilon_i(t) =
\text{ran}[0,1]$:
\begin{equation}
\sigma_i(t+1) = 
\begin{cases}
1 & \text{if $\Upsilon_i(t) < \glc 1+ 
                  \expa{-2 \beta h_i(t)} \grc ^{-1}$}\\
-1 & \text{else, }
\end{cases}, 
\elabel{heat_bath}
\end{equation}
where $h_i(t)= \sum_j J_{ij} \sigma_j(t)$ is the local field. The
square lattice is bipartite. This allows us to update one entire
sublattice simultaneously, in one \quot{sweep}, from the spins on the
other sublattice, using a vector  of random numbers $\Upsilonvec(t)=
\SET{\Upsilon_1(t) \TO \Upsilon_N(t)}$. Two subsequent sweeps update the
whole lattice.  In the remainder of the Letter, we explain our algorithm
for a unique instance of the two-dimensional Ising spin glass, defined
by a specific choice of $ \SET{J_{ij}=\pm 1}$ for nearest neighbors $i$
and $j$ on the $64 \times 64$ square lattice with periodic boundary
conditions.  We note that, in principle,  any configuration $\sigmavec$
is described through one sublattice, that is, on $32 \times 64$ sites.

In the \quot{coupling from the past} approach,  one considers
the simulation as running between an initial configuration,
at $ t = - \infty$, and the present configuration, at $t=0$ (see
\fig{coupling_from_the_past}).  The configuration $\sigmavec(t=0)$ is
an exact sample, because it results from an infinitely long Monte Carlo
calculation. It is evidently impossible to perform an infinitely long
simulation, but we may pick it up at an intermediate time, $t = t_0 <
0 $,  where, in principle, the Markov chain could be in any one of the
$2^N$ configurations, and determine $\sigmavec(t=0)$, if the Markov chain
"couples" between $t=t_0$ and $t=0$.  This means in our context  that
under the dynamics of \eq{heat_bath} all the $2^N$ possible initial
configurations $\sigmavec(t=t_0)$ yield the same configuration after a
finite number $\taucoup$ of sweeps.  If the chain does not couple, we need
to complement the sequence of random numbers $\SET{\Upsilonvec(t_0) \TO
\Upsilonvec(-1)}$ by values corresponding to earlier times.  The coupling
from the past approach to exact sampling relies on the generic property
of Markov chains to couple during their evolution \cite{propp}.

\begin{figure}[htbp]
   \centerline{
   \epsfxsize=6.8cm
   \epsfbox{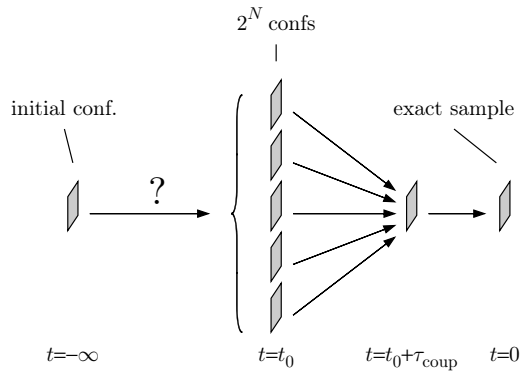}}
   \caption{Schematic representation of coupling from the past: The
   spin configuration $\sigmavec(t=0)$ is completely decorrelated
   from the initial configuration at $t= -\infty$.  The configuration
   $\sigmavec(0)$ follows from the $2^N$ configurations at $t=
   t_0$ if the chain couples between $t=t_0$ and $t=0$.}
   \flabel{coupling_from_the_past}
\end{figure}

A lower bound on the \quot{coupling time} $\taucoup$ is obtained by
checking that several randomly chosen initial configurations have
evolved toward the same state starting from time $t_0$.
\begin{figure}[htbp]
   \centerline{
   \epsfbox{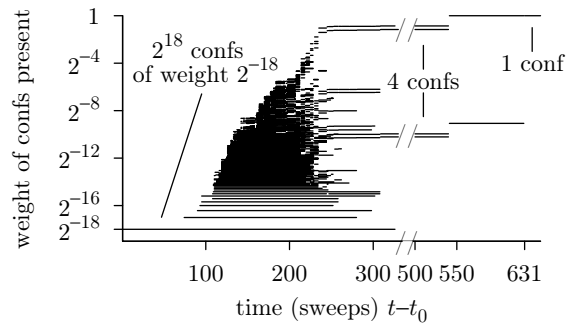}}
   \caption{Weights \emph{vs} time (in sweeps) of the heat bath algorithm
   for an instance of the $64 \times 64$ Ising spin glass at $\beta=0.5$
   (configurations that couple add weights).  At $t-t_0=631$, the $2^{18}$
   randomly chosen initial configurations have coupled toward a single
   configuration. Up to time $t\simeq t_0+330$, many
   configurations with very small weights subsist.}
   \flabel{ultra_naive}
\end{figure}
We implement this procedure in a naive algorithm which yield a
rigorous value of $\taucoup$ only if the initial configuration comprise
the entire configuration space.  We apply the naive algorithm for the
present instance of the $64\times 64$ Ising spin glass with random numbers
$\SET{\Upsilonvec(t_0), \Upsilonvec(t_0+1), \dots}$, for $2^{18}$ initial
configurations randomly chosen among the $2^{64\times 32} \simeq 3.23
\times 10^{616}$ configurations of the entire configuration space. Two
different configurations $\sigmavec(t)$ and $\tilde{\sigmavec}(t)$ may
coalesce at time $t+1$ and will remain the same from then on (see also
\cite{Derrida}), so that the number of configurations $\sigmavec(t)$
decreases with time. We may define the \quot{weight} of a configuration
$\sigmavec(t)$ as the fraction of the original configurations at time
$t_0$ that have evolved towards $\sigmavec(t)$ (see \fig{ultra_naive}). In
our example, the weight rapidly concentrates on a few configurations,
and on a single one after $t-t_0=631\le \taucoup $.  All weights $
\gg 2^{-18}$ in \fig{ultra_naive} are expected to remain essentially
unchanged for other choices for  the $\sigmavec(t_0)$ and even for
an ideal simulation taking into account all the possible initial
configurations.  At a difference with our renormalization group approach,
the naive algorithm cannot yield a rigorous upper bound for $\taucoup$
and cannot prove that, indeed, $\taucoup = 631$. We note that the coupling
time depends on the realization of the Markov chain, that is, all the
vectors $\SET{\Upsilonvec(t_0)  \TO \Upsilonvec(t_0 + \taucoup - 1)}$.

We now calculate an upper bound for $\taucoup$ (with given  random
numbers $\SET{\Upsilonvec(t_0), \dots}$).  The practical generation of
exact samples is straightforward\cite{Wilson_2,Fill}.

As shown in \fig{patches}, each spin configuration on the entire
lattice can be broken up into configurations on patches $k=1
\TO N$, that is, pieces of the lattice.  We define a set $S_k(t)$ of
\quot{configurations on patch $k$} and the product
\begin{multline}
\Omega(t) = S_1(t) \otimes S_2(t) \otimes \cdots \\
\otimes S_N(t)/\text{(pairwise compatible)}.
\elabel{Omega}
\end{multline}
At the initial time $t_0$, the set $S_k(t_0)$ contains all possible spin
configurations on the patch $k$ so that $\Omega(t_0)$ contains a superset
of the $2^N$ configurations on the entire lattice. At later times $t$,
the $S_k(t)$ contains a superset of all configurations $\sigmavec(t)$,
restricted to patch $k$.  As indicated in \eq{Omega}, the sets are
\quot{pairwise compatible}. Pairwise compatibility of $S_k(t)$ and $S_l(t)$
means that any configuration on patch $k$ must agree in the overlap
region $k \cap l$ with at least one configuration (as in \fig{patches})
of $S_l(t)$ on the neighboring patch $l$.  Pairwise compatibility is
easy to enforce by a procedure we call \quot{pruning} \cite{chanal_2}. It
consists in eliminating all spin configurations in $S_k(t)$ that lack a
compatible configuration in $S_l(t)$.  However, pairwise compatibility is
far from sufficient to assure that all elements in $\Omega(t)$ are valid
spin configurations.  We may \quot{assemble} two patches $k$ and $l$
by constructing configurations on the patch $k \cup l$ from each pair
of compatible configurations of neighboring sets $S_k(t)$ and $S_l(t)$.
In practice,  we can assemble all the patches into configurations on the
entire lattice only if the sets $S_i(t)$ are sufficiently small.

\begin{figure}[htbp]
   \centerline{
   \epsfbox{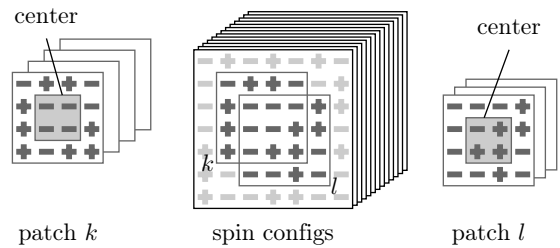}}
   \caption{A spin configuration $\sigmavec = \SET{\sigma_1 \TO
   \sigma_N}$, and the set of configurations $S_k$ and $S_l$ on patches
   $k$ and $l$. The overlap region $k \cap l$ consists here of nine sites.
   The configuration on the center of a patch,
   after one sweep, depends on the patch alone.}

   \flabel{patches}
\end{figure}

The Monte Carlo algorithm updates a spin on site $i$ as a function of
its nearest neighbors. It follows that a spin configuration in $S_k(t)$
(the set on patch $k$, of size $m\times m$, at time $t$) allows us to
determine the spin configuration at time $t+1$ for all the spins on
the \quot{center} of the patch, all the sites that do not touch its
boundary. In the example of \fig{patches}, the patches are of size
$4\times 4$, and their centers of size $2 \times 2$.  More generally,
the center of an $m\times m$ patch is of size $(m-2)\times (m-2)$.

Many different configurations of $S_k(t)$ yield, after one sweep,
identical center configurations.  The different configurations on
nine neighboring $(m-2)\times (m-2)$ centers can be assembled to form,
after pruning, the sets $S_k(t+1)$, for $k=1 \TO N$. The set $S_k(t+1)$
remains a superset of all configurations $\sigmavec(t+1)$, restricted
to patch $k$. Because of the coupling property of the heat bath algorithm,
the size of the sets $S_k(t+1)$ for small values of $\tau= t-t_0$,
is generally smaller than the size of $S_k(t)$. After many sweeps,
the coupling property is offset by the loss of information during the
assembly of the centers into the $m\times m$ patches, and the size of
the sets $S_k(t)$ starts to fluctuate around a constant value. At this
point it becomes convenient to assemble patches of size $(m+2)\times
(m+2)$, and to apply the above procedure to these larger patches. In
this renormalization procedure, the small length scales are effectively
\quot{integrated out} by the coupling property of the Monte Carlo
algorithm, and one is able to construct the configurations $\sigmavec(t)$
on ever increasing length scales.
The pruning and assembly of patches has been implemented in the \textsc{PERL}
programming language using hashing tables and referencing procedures.
The programs also transmits pairwise compatibilities from time $t$
to $t+1$.  This is highly efficient during the assembly.

\begin{figure}[htbp]
   \centerline{
   \epsfbox{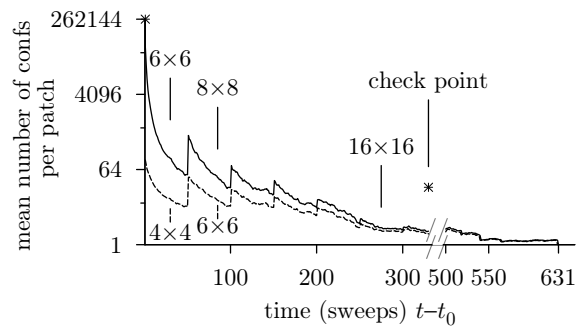}}
   \caption{Coupling of the $2^{64\times32}$ configurations of the
   two-dimensional Ising spin glass on a $64\times64$ lattice at
   $\beta=0.5$ followed through patches of increasing size (compare with
   \fig{ultra_naive}). }
   \flabel{renormalization}
\end{figure}
We now apply the renormalization-group algorithm for determining
a rigorous upper bound for $\taucoup$ to the same instance of the
$64\times 64$ Ising spin glass that was already followed with the
naive algorithm in \fig{ultra_naive}.  We follow the evolution of the
heat bath algorithm through patches and centers of size $6 \times 6$
and $4 \times 4$, respectively, starting with all the $2^{18}=262144$
possible configurations on each $6\times 6$ patch at the initial time
$t=t_0$.  We increase the size of the patches after each  $50$ sweeps
(see \fig{renormalization}).  At later times, we are able to assemble
all configurations on the $64\times 64$ lattice from the patches. At time
$t-t_0=330$, for example, the patches are of size $18\times 18$, and the
complete assembly yields $24$ configurations $\sigmavec(t)$ of the $64
\times 64$ system (see the \quot{check point} in \fig{renormalization}).
As a consistency check of our algorithm, we verified that these $24$
configurations include the $4$ configurations present in the naive
algorithm at this point (see \fig{ultra_naive}).  The algorithm couples
at time $t-t_0=631$, that is, it reaches one configuration per patch.
During the whole procedure, no configuration was dropped, and therefore,
necessarily, the complete assembly of these patches (of size $30 \times
30$) yields a single configuration $\sigmavec(t_0+631)$.  In this example,
we find that the upper bound for $\taucoup$ agrees with the lower bound
from \fig{ultra_naive}.

To conclude our discussion of the instance of the $64\times 64$ Ising spin
glass, we note that our renormalization approach has coupled after the
same number of sweeps as the straightforward Monte Carlo calculation of
the naive algorithm in \fig{ultra_naive}.  In this sense, the description
of spin configurations through the overcomplete set $\Omega(t)$ of
\eq{Omega} is efficient, even though all the $2^{32\times 64} \sim
10^{616}$ configurations that comprise the entire state space of the
system have been monitored.  However, although both simulations have
converged in $631$ sweeps of \emph{physical} time, it should be clear
that our renormalization algorithm remains very costly  in CPU time
(for details see \cite{chanal_2}). This was the price to pay in order
to exhibit an exact sample.

We have tested  our procedure for the Ising spin glass at lower
temperatures. At $\beta=0.60$, we checked on many samples of the $J_{ij}$
that we can still couple the Ising spin glass without problems. The number
of configurations  per $m \times m$ patch levels off after awhile, and the
renormalization procedure of \fig{renormalization} becomes crucial. We
have also successfully tested the procedure in the ferromagnetic Ising
model below the Curie temperature. In analyzing the behavior of our
algorithm, the naive algorithm of \fig{ultra_naive} (which generated a
sharp lower bound in our spin glass example) becomes a valuable tool.

In conclusion, we have presented a renormalization group approach to
exact sampling, and applied it to a large instance of the two-dimensional
Ising spin glass.  Our approach allows us to control a huge number of
configurations by means of patches whose sizes increase during the
simulation.

Our method can be expected to work generally for models with local
interactions but does not rely on special properties, as the partial
ordering of configurations, that only hold for severely restricted
classes of models. The fact that the model is two-dimensional only
simplifies the assembly of local information about the configurations
(the patches) but is not required to make it work. We have successfully
implemented the approach for the three-dimensional Ising spin glass
\cite{chanal_2} at temperatures lower than can be handled with a previous
method \cite{huber,Childs}.

It would be most exciting if exact sampling could now be applied to even
more complex problems, as to the three-dimensional spin glasses around
the transition temperature, where the question of whether a Monte Carlo
simulation has converged is especially difficult to answer. As mentioned
throughout this Letter, the \quot{coupling from the past} approach allows one
to draw exact samples (that are proven to have converged) and it puts all
Monte Carlo simulations for which it can be applied on an extremely solid
foundation. In this Letter we hope to have made a crucial step towards
the application of exact sampling to physically challenging problems.

\end{document}